# Analysis of the current status of tuberculosis transmission in China based on a heterogeneity model

Chuanqing Xu[1*], Kedeng Cheng[1], Yu Wang[1], Songbai Guo[1], Maoxing Liu[1], Xiaojing Wang[1], Zhiguo Zhang[2]

1 School of Science, Beijing University of Civil Engineering and Architecture, 100044, China

2 Beijing Changping District Tuberculosis Control Center, 102202, China

**Abstract:** Tuberculosis (TB) is an infectious disease transmitted through the respiratory system. China is one of the countries with a high burden of TB. Since 2004, an average of more than 800,000 cases of active TB have been reported each year in China. Analyzing the case data from 2004-2018, we find significant differences in TB incidence by age group. Therefore, the effect of age heterogeneous structure on TB transmission needs further study. We develop a model of TB to explore the role of age heterogeneity as a factor in TB transmission. The model is fitted numerically using the nonlinear least squares method to obtain the key parameters in the model, and the basic reproduction number $R_v \approx 0.8017$ is calculated and the sensitivity analysis of $R_v$ to the parameters is given. The simulation results show that reducing the number of new infections in the elderly population and increasing the recovery rate of elderly patients with the disease could significantly reduce the transmission of tuberculosis. Furthermore the feasibility of achieving the goals of the WHO End TB Strategy in China is assessed, and we obtain that with existing TB control measures it will take another 30 years for China to reach the WHO goal to reduce 90% of the number of new cases by year 2049. However, in theoretical it is feasible to reach the WHO strategic goal of ending tuberculosis by 2035 if the group contact rate in the elderly population can be reduced though it is difficulty to reduce the contact rate.

**Keywords:** tuberculosis；age heterogeneity；basic reproduction number；sensitivity analysis；partial rank correlation coefficient analysis

## 1. Introduction

Tuberculosis is an ancient disease with a worldwide distribution and is the leading cause of death from bacterial infections. Mycobacterium tuberculosis, commonly known as Mycobacterium tuberculosis, is the causative agent of tuberculosis. It was discovered and proved to be the causative agent of human tuberculosis by the German bacteriologist Koch in 1882, and it can invade all organs of the body, but is most common in causing pulmonary tuberculosis. Tuberculosis is mainly transmitted through the respiratory tract, and the source of infection is contact with TB patients who have excreted the bacteria[1].In the 19th century, tuberculosis became a major epidemic in Europe and elsewhere, spreading to all levels of society and causing one out of every seven deaths from tuberculosis, known as the "Great White Plague"[2].Due to the low effectiveness of drugs used to treat tuberculosis, the disease is uncontrollable and remains widespread worldwide. Between 1993 and 1996, the number of TB cases worldwide increased by 13%, and TB killed more people than AIDS and malaria combined. In late 1995, the World Health Organization (WHO) established March 24 as World TB Day to further promote global awareness of TB prevention and control [3]. Approximately 80% of new TB cases worldwide occur in 22 high-burden countries, with India and China accounting for the largest number of cases, 26% and 12% of global cases, respectively [4].To this day, TB remains the leading cause of disease and death in most high-incidence countries [3]. To end the global TB epidemic, WHO proposed in 2014 a post-2015 global end-tuberculosis strategy target of a 50% reduction in TB incidence by 2025 (compared to 2015) and a 90% reduction in new cases by 2035.

* corresponding author, xuchuanqing@bucea.edu.cn

Mathematical models have become a powerful tool for analyzing epidemiological characteristics [5].Many scholars developed mathematical models reflecting to the characteristics of tuberculosis based on its transmission mechanism, principles of biology, seasonal characteristics and social influences. In 1962, Waaler developed the first model of tuberculosis transmission kinetics [6].In 1967 Brogger further refined Waaler's model. He not only introduced heterogeneity but also changed the method of calculating the incidence rate, but did not give the relationship between infection and incidence rates [7]. ReVelle developed the first nonlinear differential equation model for tuberculosis using Brogger and Waaler's model as a template [8].E. Ziv et al. studied the effect of early treatment on the incidence of tuberculosis and found that early treatment reduced the incidence of tuberculosis if the treatment rate for active tuberculosis was increased from 50% to 60%. Carlos Castillo-Chavez et al. studied the role of mobility and health disparities on the transmission dynamics of Tuberculosis [9].In addition, medical studies have shown that anti-tuberculosis drugs can reduce the length of treatment for tuberculosis [10].Meanwhile, many studies have considered the effects of drug-resistant cases [11], time lag [12] and age structure [13]. However, few studies use models of TB with different age groupings; therefore, based on the collected data and the observed data characteristics (Figure 2), a susceptible-exposed-infectious-recovered (SEIR) model with different age groups is developed and the feature of heterogeneity is considered in the model to assess the effect of age as a factor on TB transmission.

The main research of this paper is as follows: in Section 2, the detailed data collected are given, and the data are analyzed in relation; a kinetic transmission model is developed, the main parameters in the model are fitted, and the value of the basic reproduction number $R_v$ is calculated. In Section 3, a sensitivity analysis of the basic reproduction number $R_v$ is performed, considering the effect of the proportion of preferential exposure within the group on $R_v$; finally, a feasibility assessment of the WHO strategic objective of ending TB is presented. Section 4 contains the discussion section.

## 2. Materials and Methods

### 2.1. Data Analysis

China is one of the high TB burden countries and faces a serious TB epidemic. The burden of TB in China has increased in the last two decades due to the emergence of drug-resistant strains of Mycobacterium tuberculosis [14], with an average of more than 800,000 new infections per year, and the number of new cases from 2004-2021 is in Figure 1(A), which shows that the number of new infections per year is decreasing year by year; a three-dimensional plot of the incidence by age group is given, and it is shows that the incidence of tuberculosis in different age groups is significantly different. The incidence rate was lowest in the 0-15 age group and much higher in the 60+ age group than in the other age groups; the number of TB cases was highest in the 20-25 and 60-65 age groups, the results of which are shown in Figure 1(B, C).Statistical data on the morbidity of tuberculosis in each age group are shown in Figure 1(D).

According to the data of the United Nations [15], the rate of aging in China is gradually accelerating, and the incidence of tuberculosis in different age groups was counted, mainly including the mean number of tuberculosis incidence, the mean incidence rate and its 95% confidence interval(CI) for each age group in the past fifteen years, and the results of statistical analysis are shown in Table 1.The results show that with the gradual increase of age, the number of tuberculosis incidence shows a trend of rising and then decreasing, while the incidence of TB showed a trend of increasing.

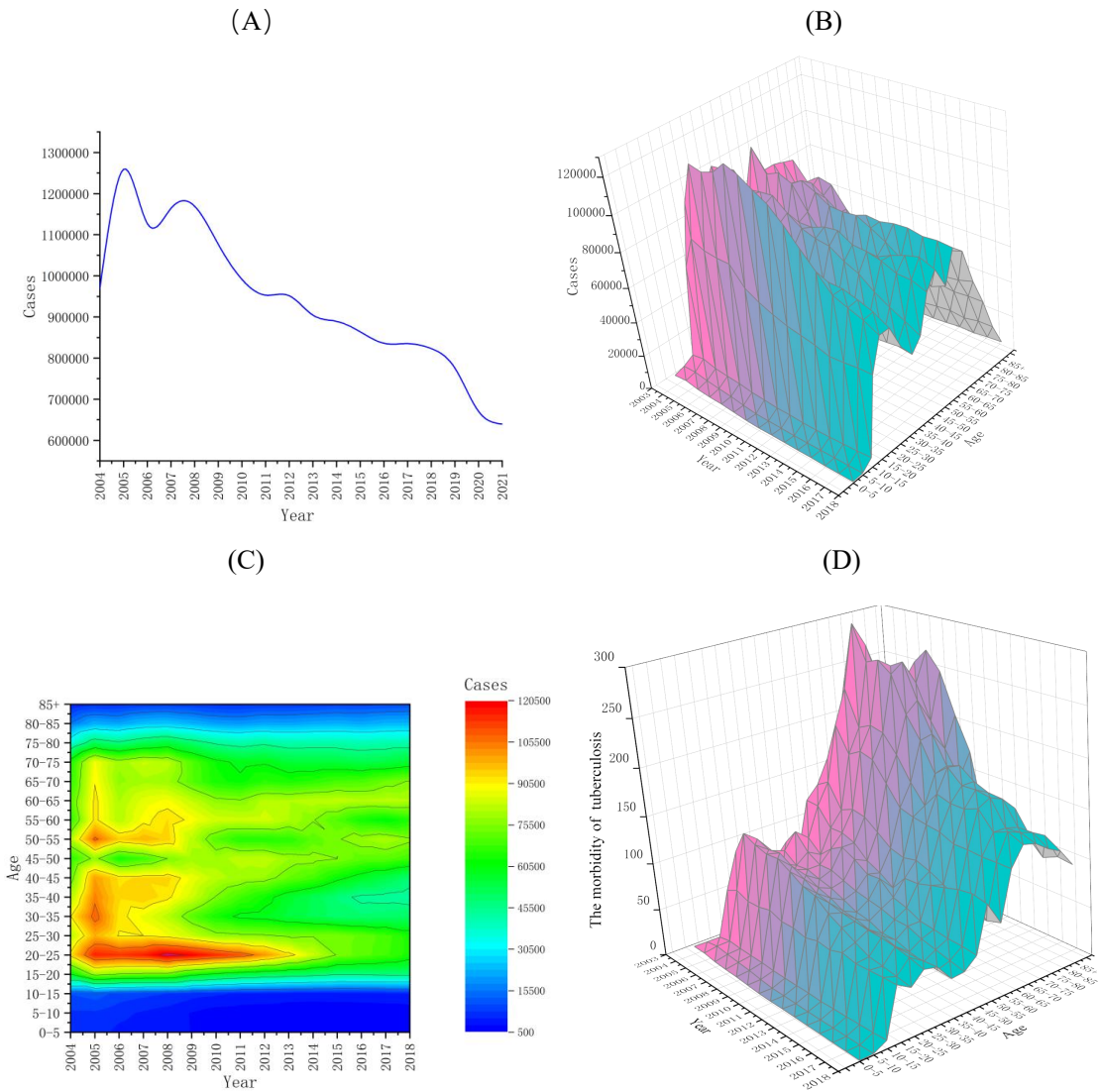

**Figure 1.** Confirmed cases of tuberculosis in mainland China in the past two decades. (A) Number of new infections per year. (B) Three-dimensional plots of TB data at different ages. (C) Contour plots of TB data at different ages. (D) Three-dimensional plot of TB morbidity data at different ages.

**Table 1.** Number and incidence of tuberculosis by age group and their confidence intervals.

| Age grouping | Mean number of incidences | 95%CI | Mean incidence rate | 95%CI |
|---|---|---|---|---|
| 0-5 | 2692 | 1645-3738 | 3.4073 | 2.0316-4.7830 |
| 5-10 | 3005 | 1619-4391 | 3.9197 | 2.1201-5.7192 |
| 10-15 | 6666 | 5284-8047 | 7.8952 | 7.1978-8.5925 |
| 15-20 | 58335 | 51904-64766 | 56.2213 | 54.8779-57.5647 |
| 20-25 | 97491 | 86176-108805 | 83.3704 | 69.7977-96.9431 |
| 25-30 | 79469 | 74453-84486 | 80.3324 | 71.4972-89.1677 |
| 30-35 | 67689 | 56899-78479 | 68.0293 | 61.3837-74.6749 |
| 35-40 | 69529 | 57746-81312 | 59.7398 | 51.3281-68.1515 |
| 40-45 | 76188 | 67365-85011 | 66.8514 | 57.3751-76.3276 |
| 45-50 | 74451 | 71112-77790 | 73.0433 | 64.3025-81.7842 |
| 50-55 | 80333 | 73185-87480 | 95.4184 | 87.3205-103.5163 |
| 55-60 | 78587 | 72486-84689 | 101.3584 | 88.3025-114.4143 |
| 60-65 | 81710 | 78840-84580 | 146.2743 | 131.4336-161.1151 |
| 65-70 | 71538 | 67286-75789 | 170.6339 | 154.6441-186.6238 |
| 70-75 | 65196 | 58412-71981 | 201.5498 | 172.0765-231.0232 |
| 75-80 | 44653 | 41831-47474 | 195.5469 | 168.0045-223.0892 |
| 80-85 | 21985 | 20706-23264 | 175.2472 | 151.4695-199.0249 |
| 85+ | 8164 | 7339-8988 | 163.7941 | 126.7235-200.8647 |

Where the incidence rate is in units of 1/100,000.

Based on the statistical data from 2004-2018, a cluster analysis of the data on the incidence of TB in different age groups is given, the complete classification process for all data is shown in Figure 2, and their complete clustering results are shown in Table 2, where distance refers to a distance between a member of each class and the center of that class. And center in this case refers to a concept similar to the intra-group average. We obtain the age group 0-15 years is the category with the lowest incidence of TB; the age group 15-60 years is the category with higher incidence; and the group over 60 years is the category with the highest incidence.

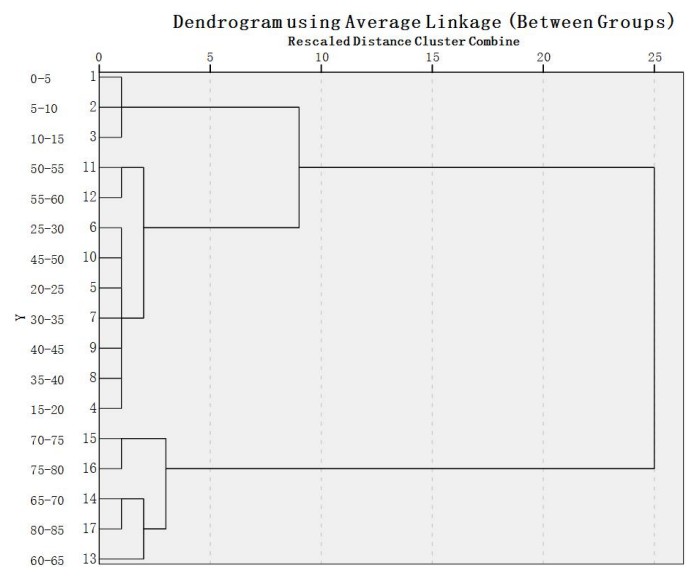

**Figure 2.** Clustered analysis of incidence data for different age groups.

**Table 2.** Clustering results by age groups based on the incidence of tuberculosis.

| Cluster Membership | | | |
|---|---|---|---|
| **Case Number** | **Age Group** | **Cluster** | **Distance** |
| 1 | 0-5 | 1 | 6.846 |
| 2 | 5-10 | 1 | 5.865 |
| 3 | 10-15 | 1 | 11.941 |
| 4 | 15-20 | 2 | 90.778 |

| 5 | 20-25 | 2 | 51.670 |
|---|---|---|---|
| 6 | 25-30 | 2 | 32.120 |
| 7 | 30-35 | 2 | 33.865 |
| 8 | 35-40 | 2 | 63.602 |
| 9 | 40-45 | 2 | 40.032 |
| 10 | 45-50 | 2 | 30.071 |
| 11 | 50-55 | 2 | 93.997 |
| 12 | 55-60 | 2 | 106.104 |
| 13 | 60-65 | 3 | 133.339 |
| 14 | 65-70 | 3 | 58.259 |
| 15 | 70-75 | 3 | 106.789 |
| 16 | 75-80 | 3 | 80.648 |
| 17 | 80-85 | 3 | 32.058 |

The third column in the table shows the final clustering results.

We analyzed the influence of age as a factor in tuberculosis infection, see Figure 3, which shows the important role of age as a factor in the transmission of tuberculosis. Therefore, in this paper, we develop a tuberculosis epidemic model that includes age-group heterogeneity, and through qualitative analysis and numerical simulation, we predict the future incidence of tuberculosis in China, and assess whether China could meet the WHO strategic target by 2035 with the current control measures, exploring measures to more effectively prevent and control tuberculosis outbreaks.

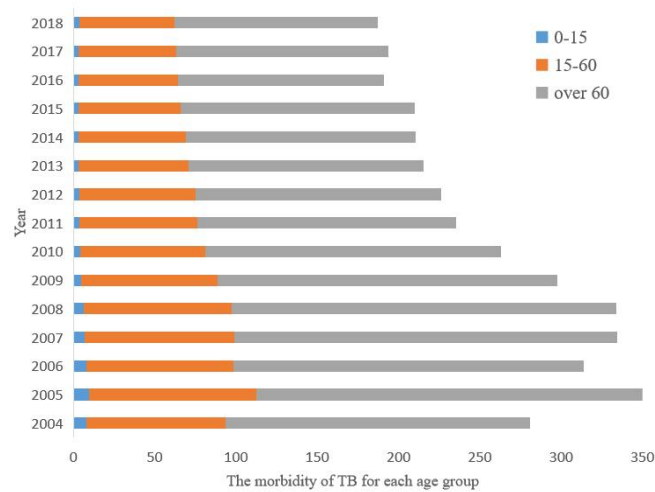

**Figure 3.** Age structure-based tuberculosis incidence data in China, 2004-2018. (Source: CDC, China [3]).

### 2.2. Model building and analysis
#### 2.2.1. Data Collection

The annual number of TB cases reported in mainland China from 2004 to 2021 was obtained from the Public Health Sciences Data Center [1]. Over a period of nearly 20, the number of reported cases exceeded 16 million. Among them, the number of cases between the ages of 15-60 years is the highest, with an average of 680,000 new infections per year, or 69%; the number of cases between the ages of 0-15 years is the lowest, with an average of 12,000 new infections per year, or

1.3%, as shown in Table 3-6. Looking at the overall TB data in China, the number of new infections per year is gradually decreasing from the initial 970,279 cases in 2004 to 639548 cases in 2021.

**Table 3.** Annual reported cases of tuberculosis infection by age group in China.

| Year | 2004 | 2005 | 2006 | 2007 | 2008 | 2009 | 2010 | 2011 | 2012 |
|---|---|---|---|---|---|---|---|---|---|
| 0-15 | 24247 | 26048 | 20735 | 17972 | 16011 | 12320 | 9751 | 8275 | 8058 |
| 15-60 | 688080 | 881944 | 795790 | 811150 | 814952 | 759035 | 705680 | 679621 | 662458 |
| 60+ | 257952 | 351316 | 311046 | 334837 | 338577 | 305583 | 275919 | 265379 | 280992 |
| Sum | 970279 | 1259308 | 1127571 | 1163959 | 1169540 | 1076938 | 991350 | 953275 | 951508 |
| Year | 2013 | 2014 | 2015 | 2016 | 2017 | 2018 | 2019 | 2020 | 2021 |
| 0-15 | 7070 | 6695 | 6861 | 6769 | 7037 | 7591 | | | |
| 15-60 | 626042 | 606179 | 575854 | 553732 | 543950 | 526626 | | | |
| 60+ | 271322 | 276507 | 281300 | 275735 | 284206 | 289125 | | | |
| Sum | 904434 | 889381 | 864015 | 836236 | 835193 | 823342 | 775764 | 670538 | 639548 |

Where data for each age group for 2019-2021 were not reported.

### 2.2.2. Model Building

In this section we develop a model of TB dynamics that includes age heterogeneity, vaccination, and divides the entire population into 3 groups according to TB incidence, with 0-15 years as the first group, 15-60 years as the second group, and over 60 years as the third group. Each group is also divided into susceptible (S), exposed (E), infectious (I), and recovered (R). Natural death and the natural death rate is different for different age groups; the age transition from the previous age group to the next age group is considered. Exposed TB cases refer to individuals who have been infected with TB bacteria but are asymptomatic, the exposed patients and reinfection of recovered individuals has no infectious in the model. The dynamic process of TB transmission is shown in Figure 4.

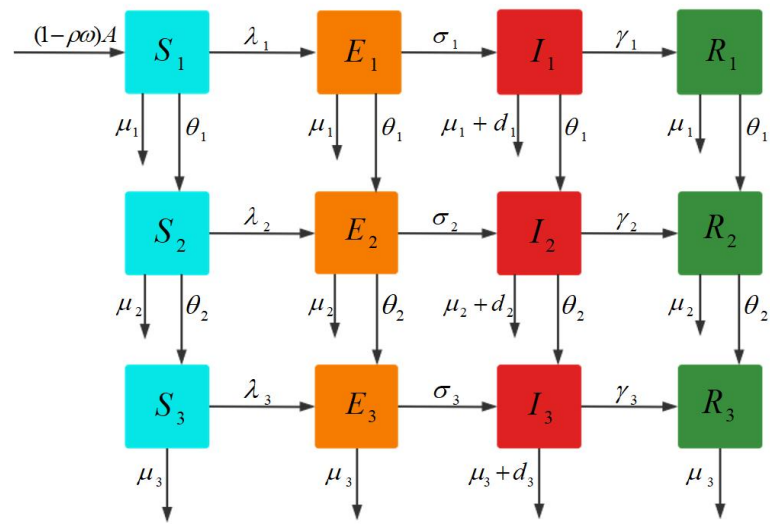

**Figure 4.** Flow chart of TB transmission with age-structure.

The model is：

$$\begin{cases} \dfrac{dS_1}{dt} = (1-\rho\omega)A - (\lambda_1 + \mu_1 + \theta_1)S_1 \\ \dfrac{dE_1}{dt} = \lambda_1 S_1 - (\sigma_1 + \mu_1 + \theta_1)E_1 \\ \dfrac{dI_1}{dt} = \sigma_1 E_1 - (\gamma_1 + d_1 + \mu_1 + \theta_1)I_1 \\ \dfrac{dR_1}{dt} = \rho\omega A + \gamma_1 I_1 - (\mu_1 + \theta_1)R_1 \\ \dfrac{dS_2}{dt} = \theta_1 S_1 - (\lambda_2 + \mu_2 + \theta_2)S_2 \\ \dfrac{dE_2}{dt} = \lambda_2 S_2 + \theta_1 E_1 - (\sigma_2 + \mu_2 + \theta_2)E_2 \\ \dfrac{dI_2}{dt} = \sigma_2 E_2 + \theta_1 I_1 - (\gamma_2 + d_2 + \mu_2 + \theta_2)I_2 \\ \dfrac{dR_2}{dt} = \gamma_2 I_2 + \theta_1 R_1 - (\mu_2 + \theta_2)R_2 \\ \dfrac{dS_3}{dt} = \theta_2 S_2 - (\lambda_3 + \mu_3)S_3 \\ \dfrac{dE_3}{dt} = \lambda_3 S_3 + \theta_2 E_2 - (\sigma_3 + \mu_3)E_3 \\ \dfrac{dI_3}{dt} = \sigma_3 E_3 + \theta_2 I_2 - (\gamma_3 + d_3 + \mu_3)I_3 \\ \dfrac{dR_3}{dt} = \gamma_3 I_3 + \theta_2 R_2 - \mu_3 R_3 \end{cases} \quad (1)$$

$A$ denotes the annual number of births in the population; $\rho$ denotes the BCG vaccination rate at birth; $\omega$ denotes the effective rate of BCG; $\mu_i$ denotes the natural mortality rate of members of group i; $\theta_i$ denotes the transfer rate of members of age group i into age group i+1 $(i=1,2)$; $\sigma_i$ denotes the activation rate (conversion rate) of latent TB patients; $\gamma_i$ denotes the recovery rate of TB patients; $d_i$ denotes the mortality rate due to TB infection in group i; and $\lambda_i$ denotes the infectivity of infected persons to susceptible persons among members of group i. The parameter $\lambda_i$ is related to the average number of contacts of group i members $a_i$, the probability that a member of group i is infected after each contact with an infected person $\beta_i$, and the proportion of contacts between members of group i and members of group j $c_{ij}$, as shown in expression (2). $c_{ij}$ was proposed by Jacquez et al [16] in 1988 to represent the proportion of contacts between members of group i and members of group j. The parameter $c_{ij}$ is related to the proportion of members of group i that are preferentially exposed to members within the same group $\varepsilon_i$ [16], the Kronecker function with a value of 1 when i = j and 0 otherwise $\delta_{ij}$, and the proportional mixing fraction $f_j$, as in the expression (3) and $f_j$ denotes the proportional mixing fraction, as in the expression (4).

$$\lambda_i = a_i \beta_i \sum_{j=1}^{3} c_{ij} \frac{I_j}{N_j}, \quad (2)$$

$$c_{ij} = \varepsilon_i \delta_{ij} + (1-\varepsilon_i)f_j, \quad (3)$$

$$f_j = (1-\varepsilon_j)a_j N_j \bigg/ \sum_{k=1}^{3}(1-\varepsilon_k)a_k N_k, \tag{4}$$

Here $N_j = S_j + E_j + I_j + R_j$.

### 3. Numerical simulation and sensitivity analysis

The outbreak of severe acute respiratory syndrome (SARS) in 2003 posed a challenge to the public health system in China, and the government, in an effort to better address public health issues, increased public health funding, revised laws regarding infectious disease control, implemented an Internet-based disease reporting system, and initiated a program to rebuild local public health facilities. These measures have facilitated TB control [17]. Complete data on TB cases are available on the official website of the Chinese Center for Disease Control and Prevention after 2004. When numerical fitting is performed, data from 2004 were used to calculate the initial values, data from 2005-2018 is used for parameter fitting, and data from 2019-2021 (for the whole country) is used to test the fit.

#### 3.1. Determination of parameters and initial values in the model

First we estimate the parameters in the model and the results are shown in Tables 7-8.

(a) From the data published in the China population statistic Yearbook 2005-2018 [18], the annual number of births of the population $A \approx 16440000$/year; the natural mortality rates of the three age groups were $\mu_1 \approx 0.0017$/year, $\mu_2 \approx 0.0023$/year, and $\mu_3 \approx 0.0367$/year ([19]); China started its immunization planning policy in 1992, and newborn infants must be vaccinated within 24 hours of birth BCG vaccine [20], thus $\rho = 1$.

(b) The average incubation period of tuberculosis is about 2 months [21], so the conversion rate of patients with latent tuberculosis is taken as $\sigma_i = 6$; the mortality rate $d_i = 0.0025$/year according to the WHO Global TB Report 2013 [4]; the recovery rate of infected cases is $\gamma_i = 0.496$/year [22]; each person will have contact with an average of 10-12 people per day [23], and we calculate the value of $a_1 = 4380$/year, and $a_2 = 3650$/year, $a_3 = 2920$/year; $\theta_1 = 0.079$/year, $\theta_2 = 0.0067$/year [19].

(c) Based on the proportion of each age group in 2005, the initial values $S_1(0)=264991621, S_2(0)=940454335, S_3(0)=99961110$ are calculated. Based on the number of people infected with TB in 2005 [1], the initial values $I_1(0)=26048, I_2(0)=881944, I_3(0)=351316$ are obtained. The percent of people with a TB bacteria infection but asymptomatic and those successfully treated for TB are 12.1% and 80%, respectively. Based on the number of people infected with TB in each age group in 2004, $E_1(0)=2934, E_2(0)=83257, E_3(0)=31212$ and $R_1(0)=19397, R_2(0)=550464, R_3(0)=206362$ are calculated.

(d) The number of cases of tuberculosis infection from 2005 to 2018 is fitted according to model (1). $\beta_1$ and $\omega$ are obtained using data from the first age group and nonlinear least squares fitting; $\beta_2$ is obtained using data from the second age group; and $\beta_3$ is fitted using data from the third age group.

#### 3.2. Data fitting results

China has taken various control measures to control TB transmission, a 5-year national plan in the 1980s, a 10-year national plan in the 1990s, and the modern TB control strategy (DOTS) introduced in the 20th century [24]. After the implementation of these short-term plans and long-term plans, TB is effectively controlled in China. The model takes into account the heterogeneity of age subgroups and exposure between age groups. Firstly, the fit is performed using nonlinear least squares under the assumption that the total population of each population group is constant (i.e., Ni was a constant) [25]. Secondly, the fit is performed under the assumption that Ni is not a constant according to the actual situation.

### 3.2.1. Total population of the 3 subgroup population $N_i$ is a constant

Assuming that the total population of the three groups remains unchanged, fitting according to model (1), the results of the fitted parameters are shown in Table 4. The results of fitting the number of TB cases in China according to the model are shown in Figure 5. It shows that the difference between the fitted curves and the actual data is very small. In addition, we performed a goodness-of-fit test on the fit results, and the goodness-of-fit coefficient was 0.904. The calculation of the goodness-of-fit coefficient can be found in Appendix A. The model fit is relatively good, which indicates that the established model is reliable and can reflect the changes of the actual data.

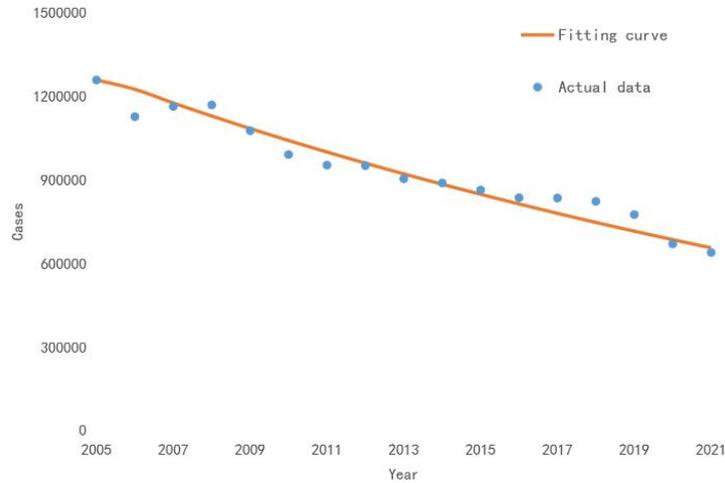

**Figure 5.** Number of new TB cases per year in mainland China and fitted curves ($N_i$ is constant). Where the blue dots **indicate** the real data and the yellow curves indicate the simulation results.

**Table 4.** Parameter values of model (1) when Ni is a constant.

| Parameter | Value | Source | Parameter | Value | Source |
|---|---|---|---|---|---|
| $A$ | 6307284 | Calculated | $\theta_2$ | 0.0039 | Calculated |
| $\rho$ | 1 | [20] | $a_1$ | 12×365 | [23] |
| $\omega$ | 0.52 | Fitting | $a_2$ | 10×365 | [23] |
| $\mu_1$ | 0.0017 | [19] | $a_3$ | 8×365 | [23] |
| $\mu_2$ | 0.0023 | [19] | $\beta_1$ | 1.157×10$^{-5}$ | Fitting |
| $\mu_3$ | 0.0367 | [19] | $\beta_2$ | 1.247×10$^{-4}$ | Fitting |
| $d_i$ | 0 | Assumption | $\beta_3$ | 3.673×10$^{-4}$ | Fitting |
| $\gamma_i$ | 0.496 | [22] | $\varepsilon_1$ | 0.4 | Assumption |
| $\sigma_i$ | 6 | [21] | $\varepsilon_2$ | 0.3 | Assumption |
| $\theta_1$ | 0.0221 | Calculated | $\varepsilon_3$ | 0.3 | Assumption |

### 3.2.2. Total population of the 3 subgroup population $N_i$ is not a constant

The numerical simulations in the previous section were performed under the condition of constant Ni, which can be considered as an ideal situation. However, in real life, the total population of each population group will change during the disease transmission process. Therefore, in this part of the model fitting process, Ni is no longer treated as a constant, and all subsequent analyses in this paper use the same treatment (i.e., Ni is no longer treated as a constant). The values of the parameters obtained from the fitting at this time are shown in Table 5 and the fitting effect is shown in Figure 6. It can be seen from the figure that the fitting results are very good, and the gap between the fitted curve and the actual data is very small. Similarly, we performed a goodness-of-fit test on the fit results and the goodness-of-fit coefficient was 0.954. The better fitting effect shows that the considered model is very trustworthy and can reflect the variation of the actual data well, even when considering more realistic situations.

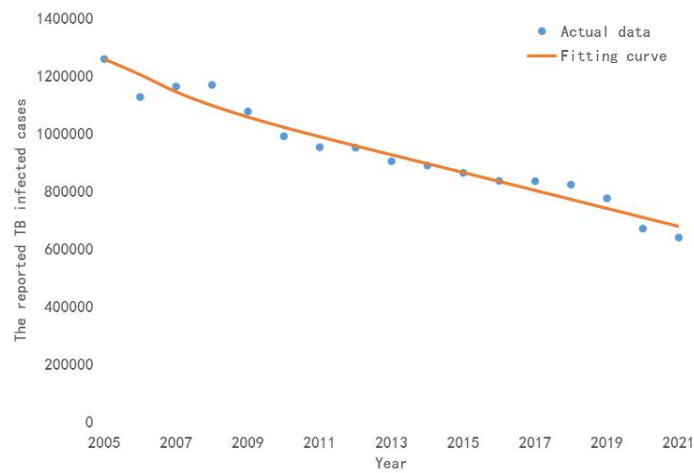

**Figure 6.** The fitting curve and observed data from 2005 to 2021 in Mainland China with Model (1) ($N_i$ is not constant). Where the blue dots indicate the real data and the yellow curves indicate the simulation results.

**Table 5.** Parameter values obtained from the model fit when Ni is not a constant

| Parameter | Value | Source | Parameter | Value | Source |
| --- | --- | --- | --- | --- | --- |
| $A$ | $1.644 \times 10^7$ | [18] | $\theta_2$ | 0.0067 | [19] |
| $\rho$ | 1 | [20] | $a_1$ | 12×365 | [23] |
| $\omega$ | 0.728 | Fitting | $a_2$ | 10×365 | [23] |
| $\mu_1$ | 0.0017 | [19] | $a_3$ | 8×365 | [23] |
| $\mu_2$ | 0.0023 | [19] | $\beta_1$ | $1.325 \times 10^{-4}$ | Fitting |
| $\mu_3$ | 0.0367 | [19] | $\beta_2$ | $7.402 \times 10^{-5}$ | Fitting |
| $d_i$ | 0.0025 | [4] | $\beta_3$ | $4.690 \times 10^{-4}$ | Fitting |
| $\gamma_i$ | 0.496 | [22] | $\varepsilon_1$ | 0.4 | Assumption |
| $\sigma_i$ | 6 | [21] | $\varepsilon_2$ | 0.3 | Assumption |
| $\theta_1$ | 0.079 | [19] | $\varepsilon_3$ | 0.3 | Assumption |

3.3. Sensitivity analysis of $R_v$

In epidemiological studies, the reproduction number (denoted as $Rv$) indicates the average number of infections in an infected person during the period of infection [26] and is one of the most important indicators to assess the risk of an infectious disease. The reproduction number $Rv$ is also considered a key epidemiological parameter in determining whether the disease can spread in an area, with $Rv > 1$ often implying that the disease will persist; $Rv < 1$ implies that the disease will become extinct. The reproduction number for model (1) was calculated using the next generation matrix method [27], and the complete calculation procedure is presented in Appendix B.

Based on the values of the parameters obtained from the fit, the basic reproduction number $Rv = 0.8017$ is calculated for model (1).

The strength of the correlation between each parameter in the model and the basic reproduction number $R_v$ is judged as a way to find the most sensitive epidemiological parameter that should be prioritized when controlling infectious diseases [28]. The sensitivity analysis of $Rv$ is performed using the partial rank correlation coefficient (PRCC) [29] of each parameter in the model, and the results are shown in Figure 7. The basic reproduction number $Rv$, $A$ (annual population births), $\beta_3$ (probability of exposure to infection in people over 60 years of age) and $\gamma_3$ (recovery rate in elderly people over 60 years of age) are the most sensitive parameters, and $|PRCC(A)| > |PRCC(\gamma_3)| > |PRCC(\beta_3)|$; Furthermore reducing new infections in the elderly population and improving recovery rate in older patients with the disease can significantly reduce the transmission of tuberculosis.

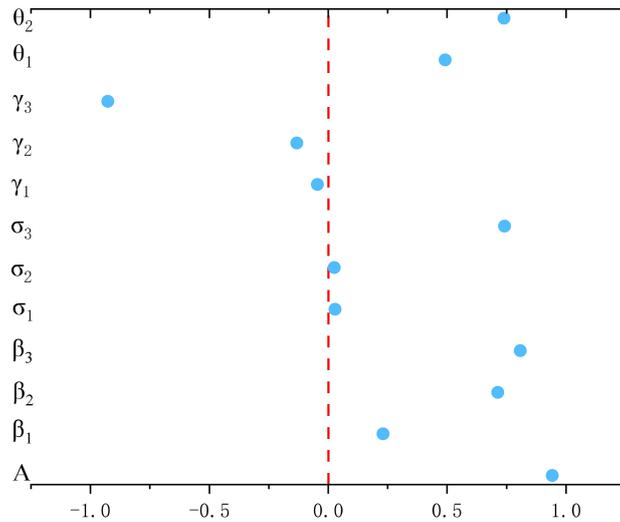

Figure 7. Plot of correlation coefficients between basic reproduction numbers and model parameters. The value of parameters are shown in Tables 5.

3.4. Effect of preferential contact proportion $\varepsilon_i$ on reproduction number $Rv$

Transmission of tuberculosis occurs mainly through close human-to-human contact, and the concept of contact is quantified in the model. Infectivity $\lambda_i$ is defined the function of parameters of the average number of contacts $a_i$, the proportion of preferential contacts within the group $\varepsilon_i$, and the probability of infection $\beta_i$. Among these parameters, the most important parameter is $\varepsilon_i$, which indicates the extent to which each individual prioritizes contacts with members of the same group. A larger $\varepsilon_i$ means that an individual has more frequent contacts with members of the same group.

Considering the effect of $\varepsilon_i$ on $Rv$ by changing the value of $\varepsilon_i$ ($i=1,2,3$) to observe the change in the basic reproduction number $R_v$. The effect of $\varepsilon_i$ on $Rv$ is considered by fixing one of the value of parameters $\varepsilon_i$ and changing the value of the others, seen Figure 8. The results show that increasing the values of $\varepsilon_1$, $\varepsilon_2$ and $\varepsilon_3$ respectively will cause an increase in the basic reproduction number $Rv$. The value of $\varepsilon_3$ has the greatest effect on $Rv$ and $\varepsilon_1$ has the least effect on $Rv$. The oldest group, whose frequent contact with each other can increase the spread of TB. Therefore, efforts to protect the elderly population should be strengthened by calling on them to increase their nutrition, exercise themselves and have regular health checks. We also call on more young people to give care to the elderly. It is believed that TB in China will be better controlled if there are fewer TB infections in the elderly group, but this is not easy to do.

(A)

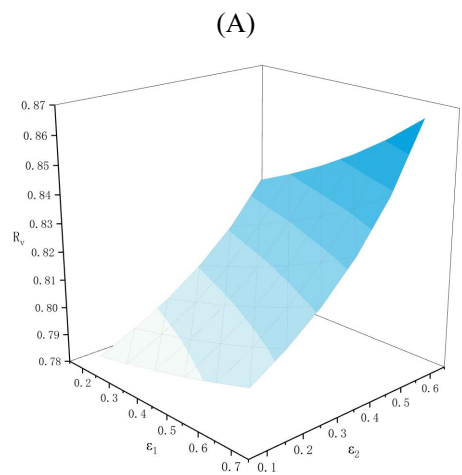

(B)

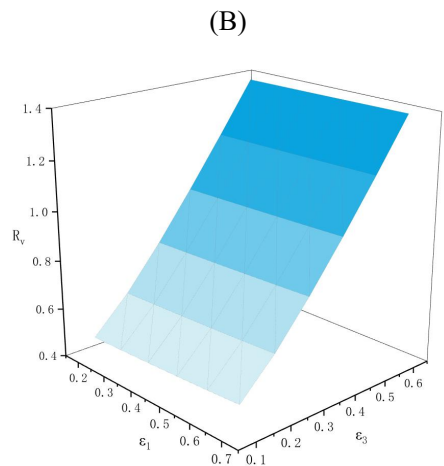

(C)

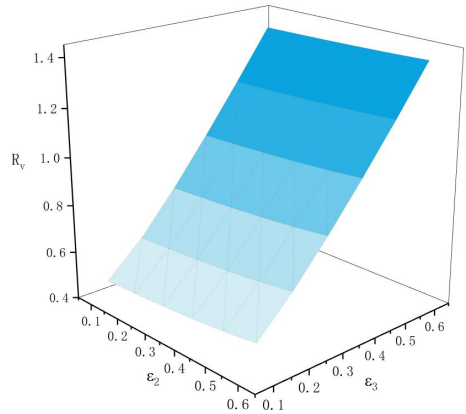

Figure 8. The comparison plot of the results of the effect of $\varepsilon_i$ on $Rv$. (A) The effect of $\varepsilon_1$ and $\varepsilon_2$ on $Rv$. (B) Effects of $\varepsilon_1$, $\varepsilon_3$ on $Rv$. (C) Effects of $\varepsilon_2$ and $\varepsilon_3$ on $Rv$.

### 3.5. Assessing the feasibility of achieving the WHO End TB Strategy in China

In the nearly two decades between 2004 and 2021, China has taken various control measures against TB and achieved very significant results, with the number of new TB cases in China decreasing from a peak of 1.25 million in 2005 to 630,000 in 2021, and significant progress has been made in controlling TB. However, globally, TB remains a huge challenge and in order to end the TB epidemic globally, WHO proposed a post-2015 global TB endgame strategy in 2014 with the strategic goal of reducing TB incidence by 50% by 2025 (compared to 2015) and new cases by 90% by 2035 [30]; and the number of new TB cases in China in 2015 was 864015, the WHO target expects China to reduce the number of new TB cases to 86402 in 2035.

In the previous analysis, we calculated that the reproduction number of tuberculosis $Rv$= 0.8017, the value of $Rv$ is less than 1, but without considering the increase of other control measures, by 2035, China will still have nearly 300,000 new infections of tuberculosis, which will not reach the desired target of WHO, the simulation results are shown in Figure 9, China will be in 2049, that is, it will take nearly 30 years to reach the desired target of WHO.

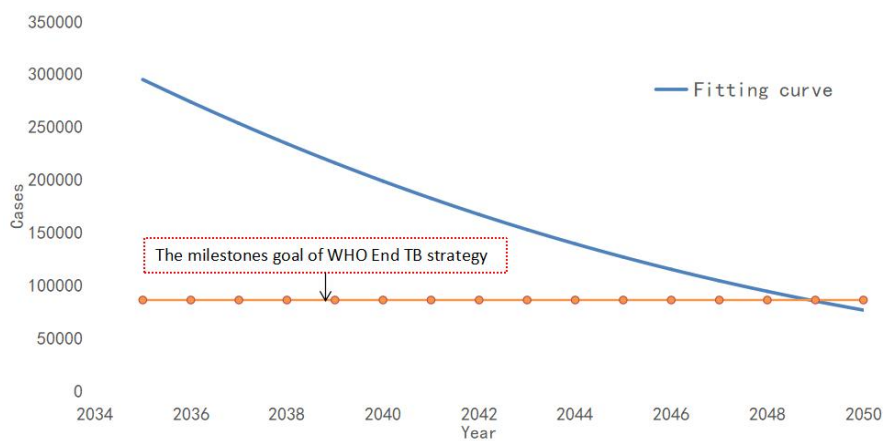

Figure 9. Projected number of TB cases after 2035 according to model (1).

### 3.6. Study of feasible control strategies

Effectiveness of BCG vaccine $\omega$, average number of contacts $a_i$ and the proportion of preferential contacts $\varepsilon_i$ are the most important factors in the TB control process. To investigate the impact of parameters $\omega$, $a_i$ and $\varepsilon_i$ on the number of new cases of tuberculosis based on current control strategies and we try to find the feasibility of achieving WHO's end-tuberculosis strategy goals. The parameter values listed in Table 5 are used to compare the control effects.

### 3.6.1. Effect of BCG vaccine effectiveness

First, considering an intervention scenario that increases the parameter $\omega$, the results are shown in Figure 10, and the number of tuberculosis cases for different values of $\omega$ is given in Figure 11. It is found that increasing the effectiveness of BCG is of limited help in reducing the number of TB cases in the short term; increasing $\omega$ by 10% would reduce the number of TB cases by an average of 3000 per year (Appendix C). Even if the effectiveness of BCG were increased from 72.8% to 95%, China would still have more than 230,000 new infections in 2035 (Appendix C), but the number of new cases per year would already be significantly reduced, so increasing the effectiveness of the vaccine would be a good option, although it would not meet the WHO strategic goal of 2035. The reason for this is that BCG is more effective in younger children but less effective in the older age groups, and the number of new cases in younger TB-infected patients have been low, so the impact of increasing the effectiveness of BCG is not significant.

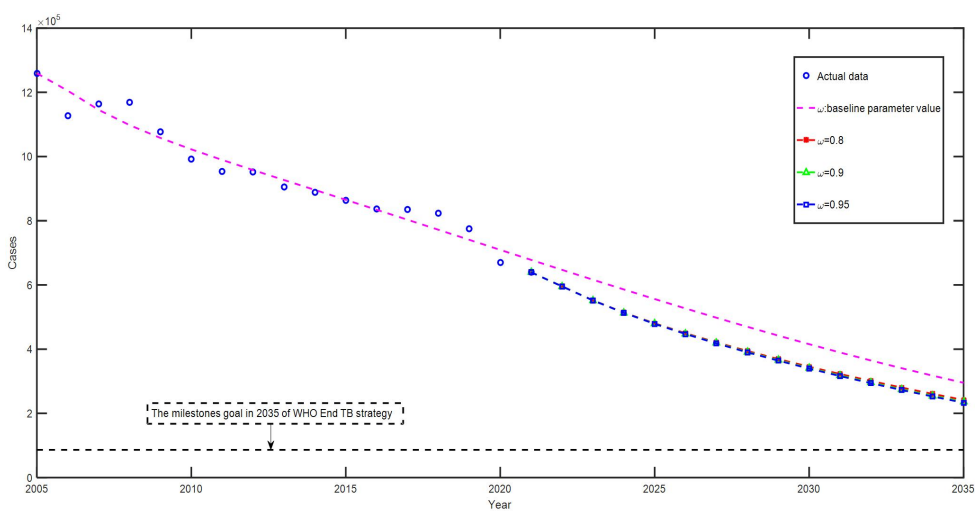

Figure 10. Effect of effectiveness $\omega$ of BCG vaccine. The blue circles indicate real data, and the black dashed lines indicate WHO strategic goals, and the other colored curves indicate different values of the parameters.

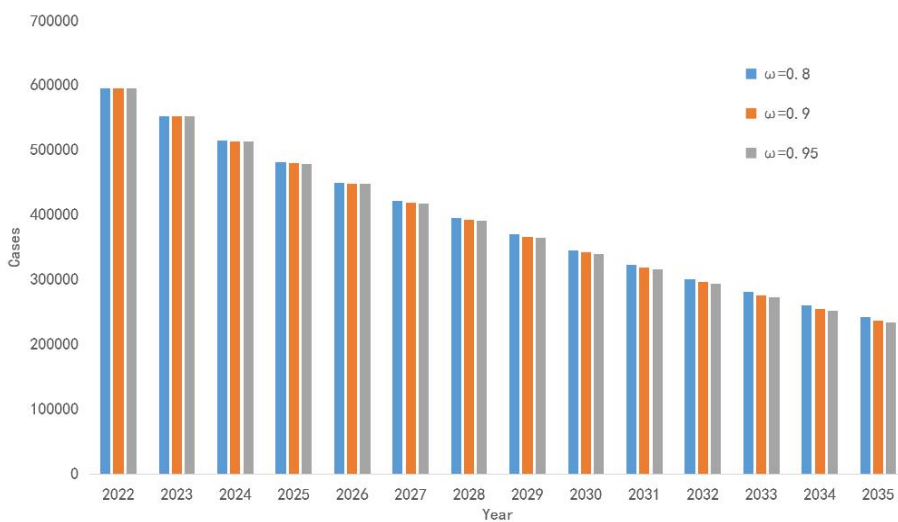

Figure 11. The number of tuberculosis cases for different values of $\omega$.

### 3.6.2. Effect of average contacts number $a_i$ on the new cases

We consider an intervention scenario that reduces the average number $a_i$ of exposures and the simulation results are shown in Figure 12. Reducing the average number of contacts $a_1$, $a_2$, and $a_3$ in each group can reduce the number of TB new cases, however reducing $a_1$ and $a_2$ by 20%, China would still have more than 200,000 new infections in 2035 and would not meet the WHO TB strategic targets. Reducing $a_1$ and $a_2$ would have lesser effect, and reducing the average number of contacts in the third group $a_3$, would have a larger effect, and reducing $a_3$ by 25% would result in a rapid decline in the number of TB infections in China. The number of new infections in 2035 is only 60,000 and the WHO target can be reached by 2035.

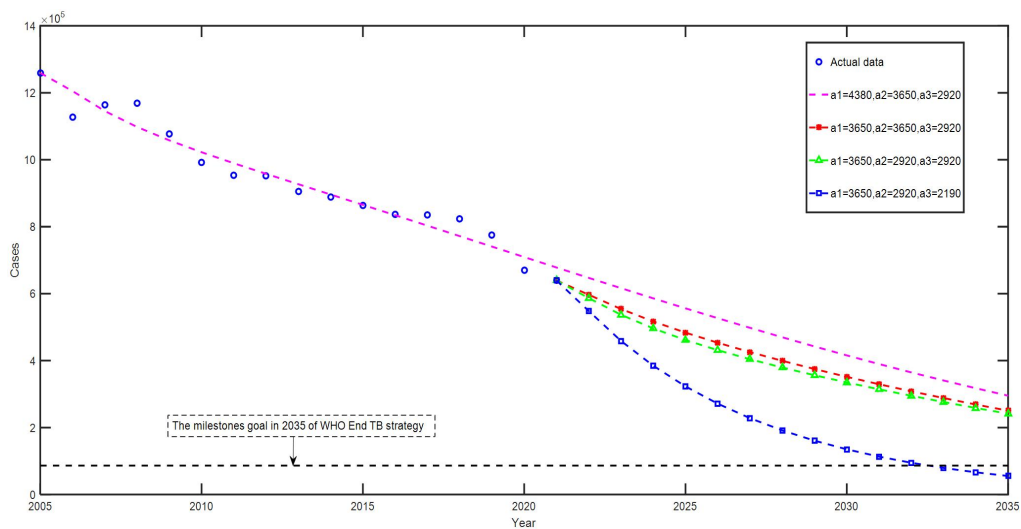

Figure 12. Effect of reducing the average number of contacts $a_i$. The blue circles are the real data, the black dashed lines are the WHO strategic targets, and the other colored curves indicate the number of new cases under different parameters.

### 3.6.3. The effect of the proportion of preferential contacts $\varepsilon_i$ within the group

Considering the impact of the proportion of preferential contacts $\varepsilon_i$ within the group, the results are given in Figure 13,14. Figure 13 illustrates that the impact caused by $\varepsilon_3$ is the largest, while Figure 14 can show that reducing only $\varepsilon_1$ or $\varepsilon_2$ has a smaller impact. It shows reducing the proportion of priority contact $\varepsilon_1$, $\varepsilon_2$, and $\varepsilon_3$ within a group can help reduce the number of new cases of tuberculosis. However the effect of reducing the parameter values $\varepsilon_1$ and $\varepsilon_2$ on the number of new cases is not very significant. Reducing $\varepsilon_1$ and $\varepsilon_2$ by 25% and 33%, respectively, would still leave China with more than 200,000 new infections in 2035 and would not meet the WHO TB strategy target. However, reducing the proportion of priority contact $\varepsilon_3$ of group 3 would have a dramatic effect. By reducing the value parameter $\varepsilon_3$ by 50%, the number of TB infections in China decreases rapidly. With only 70,000 new TB infections per year by 2035, the WHO strategic target for 2035 can be reached.

In summary for all intervention scenarios, it will be difficult for China to reach the ultimate WHO target by 2035 by using existing TB control measures. However, if the average exposure rate can be reduced by 25% or the priority exposure rate within the group can be reduced by 50%, the WHO strategic target for TB control by 2035 can be reached. China still has a long way to go on the road to eliminating tuberculosis by strengthening the implementation of tuberculosis control measures, achieving early detection and treatment, and improving the effectiveness of anti-tuberculosis drugs.

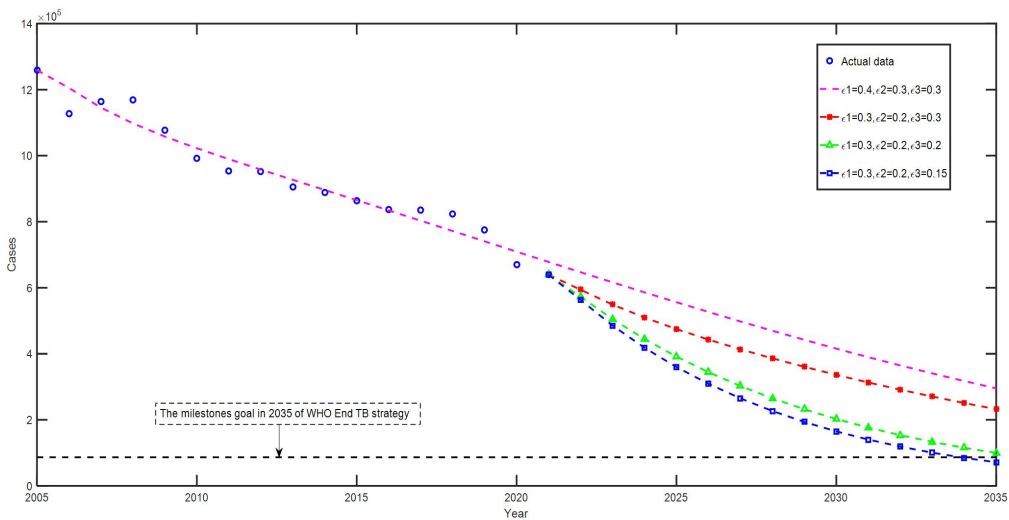

Figure 13. Effect of preferential contacts proportion $\varepsilon_i$. The blue circles are the real data, the black dashed lines are the WHO strategic targets, and the other colored curves indicate the number of new cases under different parameters.

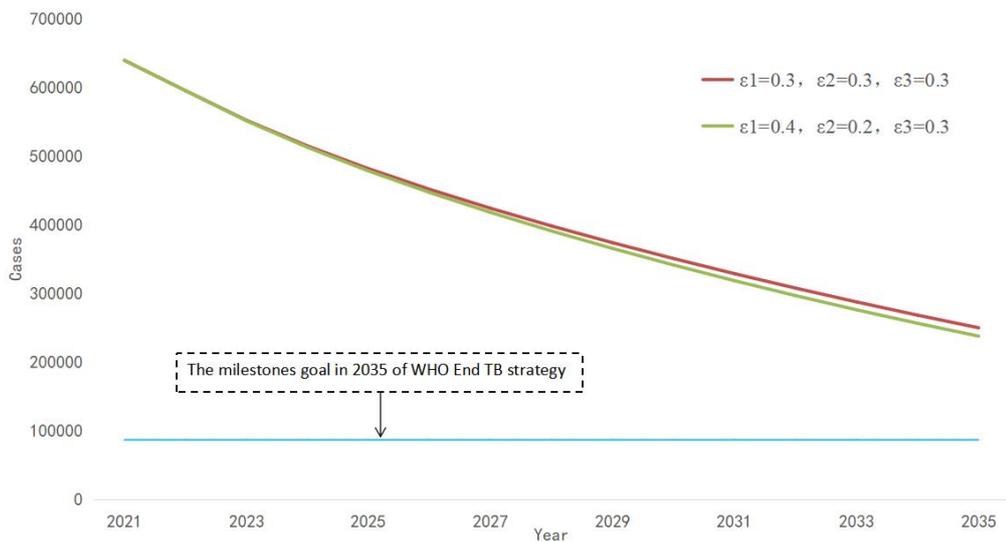

Figure 14. Assessment of the effect of lowering $\varepsilon_1$ and $\varepsilon_2$. (Compared to Figure 13, only $\varepsilon_1$ or $\varepsilon_2$ is changed)

## 4. Discussion

With TB control measures, the number of TB cases reported each year in China is gradually decreasing, which means that China is reducing the number of new cases of TB each year. But there is still a long way to go to eliminate tuberculosis, and the TB epidemic may remain a serious problem in the future. According to the data of China population Statistical Yearbook [18], China's population structure, the proportion of elderly people is increasing year by year. China considers its post-2015 end-tuberculosis strategy [31], and an aging population poses a great challenge to TB control in China. And more importantly, our study shows (Figure 3) that there were more significant differences in TB incidence between different age groups. Using TB data reported in China from 2005-2021, we develop a SEIR infectious disease model that includes three age groups: juvenile (0-15 years), middle-aged (15-60 years), and elderly (60 years or older), to investigate the role of age in the transmission of TB in China. The parameters in the model are fitted using the least squares method, and numerical simulations are performed using the fitted parameters. The fitted data are compared with the reported real data in high agreement with the annual data reported for tuberculosis in China. All of our fits were obtained from the number of reported TB cases in China, but we have no way of knowing whether the number of reported cases equals the actual number of cases, there may be underreporting, and potential underreporting may affect the estimation of some parameters and the prediction of the model, this is our deficiency. On this basis, the current basic reproduction number of TB transmission in China, $Rv \approx$ 0.8017. Even if the value of the basic reproduction number of TB transmission in China, $R_v$, is less than 1, it would still take 45 years for China to eliminate TB (to reduce the number of new infections to less than 10,000 per year). The feasibility of achieving the WHO strategic goal of ending TB by 2035 under the current TB control initiatives adopted in China is assessed, with the current control measures, it would take nearly 30 years for China to reach the expected WHO goal, and how to shorten this process is one of the issues to be considered in China.

According to the model we evaluate the effect of different intervention options, the effect of increasing the effectiveness of BCG vaccine exists, but it is limited. Even if BCG effectiveness are increased to 95%, China would not reach the WHO strategic goal of ending TB by 2035; however, if a 25% reduction in $a_3$ or a 50% reduction in $\varepsilon_3$, i.e., a reduction in overall contact in the elderly population, China could reach the WHO strategic goal by 2035. In order to eliminate TB as soon as possible, China needs to continue to strengthen the implementation of TB control measures, improve the effectiveness of TB drugs, and further explore other effective control measures.

It is reasonable to consider age grouping and contact heterogeneity in the TB model, which would be more realistic and help us to improve control strategies for TB in China [19]. Interventions such as increased nutrition for the elderly and early detection and treatment for specific groups of the elderly can be a very effective epidemic control measure [32]. Thus, our age grouping model provides a valuable foresight. For example, BCG is highly effective in young adults but less effective in middle-aged and older adults, with effectiveness in the second and third age groups being only about 50% [33]. However, with the increasing aging of the Chinese population and the fact that the elderly population has a high incidence of TB, a similar BCG control strategy should be implemented for the potentially high-risk elderly subpopulation, which may significantly reduce the incidence in this group. The constant subpopulation ($N_i$ is constant) is also better fitted in our model, and further studies of this ideal case will follow. According to reports on tuberculosis [34], approximately 0.5% to 7.2% of tuberculosis cases in developed countries are caused by Mycobacterium bovis, while in many developing countries, the severity of human infection with bovine tuberculosis is much higher than in developed countries [35]. In China, there are several regions that depend on animal husbandry, such as the pastoral areas of Xinjiang, Tibet, and Inner Mongolia, where cows are mostly raised on a small scale or free-range, which can greatly facilitate the transmission of bovine TB between humans and cattle [36]. Therefore, it is reasonable to believe that a proportion of the TB patients in China are bovine TB patients, and that bovine TB patients are more capable of transmitting the virus. If measures can be taken to control the number of infections in this group, it is believed that this will help to reduce the overall number of TB cases in China. This is an issue that we intend to continue to study. Studying the role of age in TB transmission may help to predict long-term health risks and thus suggest targeted TB control strategies, more rational programming and more efficient use of limited resources [37], which is still of great significance for TB control. Also focusing on the elderly population, improving their healthy living standards, increasing their nutrition, and calling for their greater participation in exercise all have a positive impact on TB control.

**Appendix A. Calculation of the goodness-of-fit coefficient**

We use $R^2$ to denote the goodness-of-fit coefficient, and $R^2$ is defined by the expression (5).

$$R^2 = 1 - \frac{RSS}{TSS} \tag{5}$$

Where, RSS means residual sum of squares, which represents the sum of squares of the deviations between the actual and simulated values. TSS means total sum of squares, which represents the sum of squares of the deviations between the actual and expected values.

We use the magnitude of the goodness-of-fit coefficient to judge the effectiveness of the fit. The closer the goodness-of-fit coefficient is to 1, the better the fit is.

**Appendix B. The calculation process of the basic reproduction number $R_V$**

The disease-free equilibrium point of the model (1) is $P^0 = (S_1^0, 0, 0, R_1^0, S_2^0, 0, 0, R_2^0, S_3^0, 0, 0, R_3^0)$, where

$S_1^0 = \frac{(1-\rho\omega)A}{\mu_1 + \theta_1}$, $R_1^0 = \frac{\rho\omega A}{\mu_1 + \theta_1}$, $S_2^0 = \frac{\theta_1 S_1^0}{\mu_2 + \theta_2}$, $R_2^0 = \frac{\theta_1 R_1^0}{\mu_2 + \theta_2}$, $S_3^0 = \frac{\theta_2 S_2^0}{\mu_3}$, $R_3^0 = \frac{\theta_2 R_2^0}{\mu_3}$.

The reproduction number of model (1) is calculated using the next generation matrix approach,

$$\mathcal{F} = \begin{pmatrix} a_1\beta_1 S_1(c_{11}\frac{I_1}{N_1} + c_{12}\frac{I_2}{N_2} + c_{13}\frac{I_3}{N_3}) \\ 0 \\ a_2\beta_2 S_2(c_{21}\frac{I_1}{N_1} + c_{22}\frac{I_2}{N_2} + c_{23}\frac{I_3}{N_3}) \\ 0 \\ a_3\beta_3 S_3(c_{31}\frac{I_1}{N_1} + c_{32}\frac{I_2}{N_2} + c_{33}\frac{I_3}{N_3}) \\ 0 \end{pmatrix}, \quad \mathcal{V} = \begin{pmatrix} (\mu_1 + \sigma_1 + \theta_1)E_1 \\ (\mu_1 + d_1 + \gamma_1 + \theta_1)I_1 - \sigma_1 E_1 \\ (\mu_2 + \sigma_2 + \theta_2)E_2 - \theta_1 E_1 \\ (\mu_2 + d_2 + \gamma_2 + \theta_2)I_2 - \sigma_2 E_2 - \theta_1 I_1 \\ (\mu_3 + \sigma_3)E_3 - \theta_2 E_2 \\ (\mu_3 + d_3 + \gamma_3)I_3 - \sigma_3 E_3 - \theta_2 I_2 \end{pmatrix}$$

The partial derivatives of $E_i$ and $I_i$ is obtained,

$$F = \begin{pmatrix} 0 & a_1\beta_1 c_{11} S_1^0 \frac{1}{N_1^0} & 0 & a_1\beta_1 c_{12} S_1^0 \frac{1}{N_2^0} & 0 & a_1\beta_1 c_{13} S_1^0 \frac{1}{N_3^0} \\ 0 & 0 & 0 & 0 & 0 & 0 \\ 0 & a_2\beta_2 c_{21} S_2^0 \frac{1}{N_1^0} & 0 & a_2\beta_2 c_{22} S_2^0 \frac{1}{N_2^0} & 0 & a_2\beta_2 c_{23} S_2^0 \frac{1}{N_3^0} \\ 0 & 0 & 0 & 0 & 0 & 0 \\ 0 & a_3\beta_3 c_{31} S_3^0 \frac{1}{N_1^0} & 0 & a_3\beta_3 c_{32} S_3^0 \frac{1}{N_2^0} & 0 & a_3\beta_3 c_{33} S_3^0 \frac{1}{N_3^0} \\ 0 & 0 & 0 & 0 & 0 & 0 \end{pmatrix}$$

The partial derivatives of $E_i$ and $I_i$ is obtained,

$$V = \begin{pmatrix} \mu_1+\sigma_1+\theta_1 & 0 & 0 & 0 & 0 & 0 \\ -\sigma_1 & \mu_1+d_1+\gamma_1+\theta_1 & 0 & 0 & 0 & 0 \\ -\theta_1 & 0 & \mu_2+\sigma_2+\theta_2 & 0 & 0 & 0 \\ 0 & -\theta_1 & -\sigma_2 & \mu_2+d_2+\gamma_2+\theta_2 & 0 & 0 \\ 0 & 0 & -\theta_2 & 0 & \mu_3+\sigma_3 & 0 \\ 0 & 0 & 0 & -\theta_2 & -\sigma_3 & \mu_3+d_3+\gamma_3 \end{pmatrix}$$

and

$$V^{-1} = \begin{pmatrix} \frac{1}{A_1} & 0 & 0 & 0 & 0 & 0 \\ \frac{\sigma_1}{A_1 A_2} & \frac{1}{A_2} & 0 & 0 & 0 & 0 \\ \frac{\theta_1}{A_1 A_3} & 0 & \frac{1}{A_3} & 0 & 0 & 0 \\ B_1 & \frac{\theta_1}{A_2 A_4} & \frac{\sigma_2}{A_3 A_4} & \frac{1}{A_4} & 0 & 0 \\ \frac{\theta_1 \theta_2}{A_1 A_3 A_5} & 0 & \frac{\theta_2}{A_3 A_5} & 0 & \frac{1}{A_5} & 0 \\ B_2 & \frac{\theta_1 \theta_2}{A_2 A_4 A_6} & B_3 & \frac{\theta_2}{A_4 A_6} & \frac{\sigma_3}{A_5 A_6} & \frac{1}{A_6} \end{pmatrix}$$

Among them

$A_1 = \mu_1+\sigma_1+\theta_1$, $A_2 = \mu_1+d_1+\gamma_1+\theta_1$, $A_3 = \mu_2+\sigma_2+\theta_2$, $A_4 = \mu_2+d_2+\gamma_2+\theta_2$, $A_5 = \mu_3+\sigma_3$,

$A_6 = \mu_3+d_3+\gamma_3$, $B_1 = \frac{\sigma_1 \theta_1}{A_1 A_2 A_4} + \frac{\sigma_2 \theta_1}{A_1 A_3 A_4}$, $B_2 = \frac{\sigma_1 \theta_1 \theta_2}{A_1 A_2 A_4 A_6} + \frac{\sigma_2 \theta_1 \theta_2}{A_1 A_3 A_4 A_6} + \frac{\sigma_3 \theta_1 \theta_2}{A_1 A_3 A_5 A_6}$,

$B_3 = \frac{\sigma_2 \theta_2}{A_3 A_4 A_6} + \frac{\sigma_3 \theta_2}{A_3 A_5 A_6}$

Therefore, at the disease-free equilibrium point $P^0$, there is

$$FV^{-1} = \begin{pmatrix} H & K \\ O & T \end{pmatrix},$$

Among them

$$H = \begin{pmatrix} a_1\beta_1 S_1^0(\frac{c_{11}\sigma_1}{N_1^0 A_1 A_2} + \frac{c_{12}B_1}{N_2^0} + \frac{c_{13}B_2}{N_3^0}) & a_1\beta_1 S_1^0(\frac{c_{11}}{N_1^0 A_2} + \frac{c_{12}\theta_1}{N_2^0 A_2 A_4} + \frac{c_{13}\theta_1\theta_2}{N_3^0 A_2 A_4 A_6}) & a_1\beta_1 S_1^0(\frac{c_{12}\sigma_2}{N_2^0 A_3 A_4} + \frac{c_{13}B_3}{N_3^0}) \\ 0 & 0 & 0 \\ a_2\beta_2 S_2^0(\frac{c_{21}\sigma_1}{N_1^0 A_1 A_2} + \frac{c_{22}B_1}{N_2^0} + \frac{c_{23}B_2}{N_3^0}) & a_2\beta_2 S_2^0(\frac{c_{21}}{N_1^0 A_2} + \frac{c_{22}\theta_1}{N_2^0 A_2 A_4} + \frac{c_{23}\theta_1\theta_2}{N_3^0 A_2 A_4 A_6}) & a_2\beta_2 S_2^0(\frac{c_{22}\sigma_2}{N_2^0 A_3 A_4} + \frac{c_{23}B_3}{N_3^0}) \end{pmatrix},$$

$$K = \begin{pmatrix} \frac{a_1\beta_1 c_{12} S_1^0}{N_2^0 A_4} + \frac{a_1\beta_1 c_{13}\theta_2 S_1^0}{N_3^0 A_4 A_6} & \frac{a_1\beta_1 c_{13}\sigma_3 S_1^0}{N_3^0 A_5 A_6} & \frac{a_1\beta_1 c_{13} S_1^0}{N_3^0 A_6} \\ 0 & 0 & 0 \\ \frac{a_2\beta_2 c_{22} S_2^0}{N_2^0 A_4} + \frac{a_2\beta_2 c_{23}\theta_2 S_2^0}{N_3^0 A_4 A_6} & \frac{a_2\beta_2 c_{23}\sigma_3 S_2^0}{N_3^0 A_5 A_6} & \frac{a_2\beta_2 c_{23} S_2^0}{N_3^0 A_6} \end{pmatrix}$$

$$O = \begin{pmatrix} 0 & 0 & 0 \\ a_3\beta_3 S_3^0(\frac{c_{31}\sigma_1}{N_1^0 A_1 A_2} + \frac{c_{32}B_1}{N_2^0} + \frac{c_{33}B_2}{N_3^0}) & a_3\beta_3 S_3^0(\frac{c_{31}}{N_1^0 A_2} + \frac{c_{32}\theta_1}{N_2^0 A_2 A_4} + \frac{c_{33}\theta_1\theta_2}{N_3^0 A_2 A_4 A_6}) & a_3\beta_3 S_3^0(\frac{c_{32}\sigma_2}{N_2^0 A_3 A_4} + \frac{c_{33}B_3}{N_3^0}) \\ 0 & 0 & 0 \end{pmatrix}$$

$$T = \begin{pmatrix} 0 & 0 & 0 \\ \frac{a_3\beta_3 c_{32} S_3^0}{N_2^0 A_4} + \frac{a_3\beta_3 c_{33}\theta_2 S_3^0}{N_3^0 A_4 A_6} & \frac{a_3\beta_3 c_{33}\sigma_3 S_3^0}{N_3^0 A_5 A_6} & \frac{a_3\beta_3 c_{33} S_3^0}{N_3^0 A_6} \\ 0 & 0 & 0 \end{pmatrix}$$

The reproduction number is the spectral radius of the matrix $FV^{-1}$[38], i.e. $R_v = \rho(FV^{-1})$.

# Appendix C. Additional notes to Figure 11

**Table 6.** Increasing the number of TB cases after $\omega$.

① $\omega=0.8$

| Year | 2022 | 2023 | 2024 | 2025 | 2026 | 2027 | 2028 |
|---|---|---|---|---|---|---|---|
| Cases | 595736 | 552311 | 514403 | 480536 | 449676 | 421138 | 394438 |
| Year | 2029 | 2030 | 2031 | 2032 | 2033 | 2034 | 2035 |
| Cases | 369259 | 345381 | 322660 | 300998 | 280332 | 260619 | 241832 |

② $\omega=0.9$

| Year | 2022 | 2023 | 2024 | 2025 | 2026 | 2027 | 2028 |
|---|---|---|---|---|---|---|---|
| Cases | 595663 | 551990 | 513698 | 479352 | 447964 | 418866 | 391609 |
| Year | 2029 | 2030 | 2031 | 2032 | 2033 | 2034 | 2035 |

| Cases | 365888 | 341499 | 318303 | 296209 | 275157 | 255106 | 236028 |
|---|---|---|---|---|---|---|---|

③ $\omega=0.95$

| Year | 2022 | 2023 | 2024 | 2025 | 2026 | 2027 | 2028 |
|---|---|---|---|---|---|---|---|
| Cases | 595626 | 551830 | 513347 | 478765 | 447116 | 417747 | 390221 |
| Year | 2029 | 2030 | 2031 | 2032 | 2033 | 2034 | 2035 |
| Cases | 364242 | 339610 | 316192 | 293899 | 272671 | 252468 | 233262 |

**Acknowledgments:** This study was funded by Natural Science Foundation of China (NSFC 11901027), Postgraduate Teaching Research and Quality Improvement Project of BUCEA (J2021010), BUCEA Post Graduate Innovation Project (PG2022139). We thank all the individuals who generously shared their time and materials for this study.

**References**

[1] Public Health Sciences Data Center. http://www.phsciencedata.cn

[2] Zhao, H.M. Tuberculosis Control and Health Education [J]. World's Newest Medical Information Digest, 2018,18(51):197-199.

[3] Chinese Center for Disease Control and Prevention. http://www.chinatb.org.

[4] World Health Organization. Global Tuberculosis Report; 2013. Available online: http://www.who.int/tb/publications/global_report/en/.

[5] Houben, M.G.; Wu, C.Y.; Rhines, A.S.; Denholm, J.T.; Gomez, G.B.; Hippner, P. Feasibility of achieving the 2025 WHO global tuberculosis target in South Africa, China, and India: A combined analysis of 11 mathematical models. Lancet Glob. Health 2016, 4, e806–e815.

[6] H.Waaler, A.Geser, S.Andersen. The use of mathematical models in the study of the epidemiology of tuberculosis. American Journal of Public Health.1962,52(6): 1002~1013.

[7] S.Brogger. Systems analysis in tuberculosis control: a model. American Review of Respiratory Disease.1967,95(3):419~434.

[8] C.S.Revelle, W.R.Lynn, F.Feldmann. Mathematical model for the economical location of tuberculosis control activities in developing nations. American Review of Respiratory Disease.1967,96:893~909.

[9] Moreno V, Espinoza B, Barley K, Paredes M, Bichara D, Mubayi A, Castillo-Chavez C. The role of mobility and health disparities on the transmission dynamics of Tuberculosis. Theor Biol Med Model. 2017 Jan 28;14(1):3.

[10] Koch A, Cox H, Mizrahi V. Drug-resistant tuberculosis: challenges and opportunities for diagnosis and treatment[J]. Curr Opin Pharmacol, 2018, 42:7-15.

[11] Y.J.Lin,C.M. Liao.Seasonal dynamics of tuberculosis epidemics and implications for multidrug-resistant infection risk assessment. Epidemiology and Infection. 2014,142(2):358~370.

[12] H.X.Wang, L.Q.Jiang, G.H.Wang. Stability of a tuberculosis model with a time delay in transmission. Journal of North University of China.2014,35(3):238~242.

[13] T.Cohen, M.Lipsitch, R.P.Walensky, et al. Beneficial and perverse effects of isoniazid preventive therapy for latent tuberculosis infection in HIV-tuberculosis co-infected populations. Proceedings of the National of Science of the United of America. 2006,103(18):7042~7047.

[14] Wang, Y. The fifth national tuberculosis epidemiological survey in 2010. Chin. J. Autituberc. 2012, 8, 485–508.

[15] Smith, J.P.; Strauss, S.; Zhao, Y.H. Healthy Aging in China. J. Econ. Ageing 2014, 4, 37–43.

[16] J. A. Jacquez, C. P. Simon, J. Koopman, et al. Modeling and analyzing HIV transmission: the effect of contact patterns. Mathematical Biosciences, 1988, 92(2):119-199.

[17] Millet, J.P.; Shaw, E.; Orcau, A.; Casals, M.; Miro, J.M.; Cayla, J.A. The Barcelona Tuberculosis Recurrence Working Group 'Tuberculosis Recurrence after Completion Treatment in a European City: Reinfection or Relapse? PLoS ONE 2013, 8, e64898.

[18] China Population Statistic Yearbook. 2021. Available online: http://www.stats.gov.cn/tjsj/ndsj/.

[19] Zhao Y, Li M, Yuan S. Analysis of Transmission and Control of Tuberculosis in Mainland China, 2005–2016, Based on the Age-Structure Mathematical Model. International Journal of Environmental Research and Public Health. 2017; 14(10):1192.

[20] Wang J. Can BCG vaccine protect you for life [J]. Family Medicine (second half of the month), 2015(03):14.

[21] National Scientific Data Sharing Platform for Population and Health. 2021. Available online: http://www.ncmi.cn/info/69/1544.

[22] Blower, S.M.; McLean, A.R.; Porco, T.C. The intrinsic transmission dynamics of tuberculosis epidemics. Nat. Med. 1995, 8, 815–821.

[23] J. M. Read, J. Lessler, S. Riley, S. Wang, et al. Social mixing patterns in rural and urban areas of southern China. Proceedings. Biological sciences, 2014, 281(1785).

[24] Cheng S. M., Liu E. Y., Wang F., Ma Y., Zhou L., Wang L. X., Wan L. Progress in the implementation and systematic evaluation of modern tuberculosis control strategies [J]. Chinese Journal of Anti-Tuberculosis, 2012, 34(09):585-591.

[25] Zhilan Feng, John W. Glasser, Andrew N. Hill, Mikael A. Franko, Rose-Marie Garlsson, Hans Hallander, Peet Tüll, Patrick Olin, Modeling rates of infection with transient maternal antibodies and waning active immunity: Application to Bordetella pertussis in Sweden, Journal of Theoretical Biology, Volume 356,2014,123-132.

[26] Fraser, C.; Donnelly, C.A.; Cauchemez, S.; Hanage, W.P.; Van Kerkhove, M.D.; Hollingsworth, T.D.; Griffin,J.; Baggaley, R.F.; Jenkins, H.E.; Lyons, E.J.; et al. Pandemic Potential of a Strain of Influenza A (H1N1):Early Findings. Science 2009, 324, 1557–1561

[27] P.V.D Driessche, J. Watmough. Reproduction numbers and sub-threshold endemic equilibria for compartmental models of disease transmission. Mathematical Biosciences. 2002,180(1-2):29~48.

[28] Gao D，Lou Y，He D，Porco T C，Kuang Y，Chowell G，Ruan S. Prevention and control of Zika as a mosquito-borne and sexually transmitted disease: a mathematical modeling analysis. Scientific Reports, 2016, 6:28070.

[29] Yan，L.K. Application of correlation coefficient and partial correlation coefficient in correlation analysis [J]. Journal of Yunnan Institute of Finance and Trade, 2003(03):78-80.

[30] World Health Organization. WHO End TB Strategy. 2014. Available online: http://www.who.int/tb/post2015$_$strategy/en/.

[31] Huynh, G.H.; Klein, D.J.; Chin, D.P.; Wagner, B.G.; Eckhoff, P.A.; Liu, R.Z.; Wang, L.X. Tuberculosis control strategies to reach the 2035 global targets in China: The role of changing demographics and reactivation disease. BMC Med. 2015, 13, 88.

[32] Ziv, E.; Daley, C.L.; Blower, S.M. Early therapy for latent tuberculosis infection. Am. J. Epidemiol. 2001, 153, 381–385.

[33] Ziv, E.; Daley, C.L.; Blower, S.M. Potential public health impact of new tuberculosis vaccines. Emerg. Infect. Dis. 2004, 10, 1529–1535.

[34] Guo, A.Z. Control and decontamination of tuberculosis in animals. 2019 International Symposium on Human-Veterinary Diseases and China Rabies Annual Meeting [R], Hefei, China，2019.

[35] Zhang, X.Y., Sun, M.J., Fan, W.X. Transmission and control of bovine tuberculosis[J]. China Animal Quarantine, 2017, 34(07): 0-74.

[36] Chen, Y.Y., Chen, H.C., Guo, A.Z. Epidemiological characteristics of bovine tuberculosis and its prevention and control[J]. China Animal Health, 2012, 14(04): 7-9.

[37] Liu, L.J.; Zhao, X.Q.; Zhou, Y.C. A tuberculosis model with seasonality. Bull. Math. Biol. 2010, 72, 931–952.

[38] O. Diekmann, J.A. Heesterbeek, J.A. Metz. On the definition and the computation of the basic reproduction ratio $R_0$ in models for infectious in heterogeneous populations. Journal of Mathematical Biology. 1990, 28(4):365~382.